# Making a Case for Research Collaboration Between Artificial Intelligence and Operations Research Experts

Workshop Series Final Report

**April 2025**

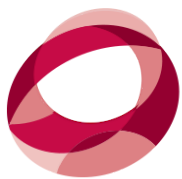

## Authors

Radhika Kulkarni, SAS Institute, Inc. (retired), rvk9@cornell.edu, https://www.informs.org/Explore/History-of-O.R.-Excellence/Miser-Harris-Presidential-Portrait-Gallery/Radhika-Kulkarni

Gianluca Brero, Bryant University, gbrero@bryant.edu, https://www.bryant.edu/academics/faculty/brero-gianluca

Yu Ding, Georgia Institute of Technology, yu.ding@isye.gatech.edu, https://www.isye.gatech.edu/users/yu-ding

Swati Gupta, Massachusetts Institute of Technology, swatig@mit.edu, https://mitsloan.mit.edu/faculty/directory/swati-gupta

Sven Koenig, University of Southern California, svenk@uci.edu, http://idm-lab.org/

Ramayya Krishnan, Carnegie Mellon University, rk2x@cmu.edu, https://www.heinz.cmu.edu/faculty-research/profiles/krishnan-ramayya

Thiago Serra, University of Iowa, thiago.serra@bucknell.edu, https://tippie.uiowa.edu/people/thiago-serra

Phebe Vayanos, University of Southern California, phebe.vayanos@usc.edu, https://viterbi.usc.edu/directory/faculty/Vayanos/Phebe

Segev Wasserkrug, IBM Research, SEGEVW@il.ibm.com, https://research.ibm.com/people/segev-wasserkrug

Holly Wiberg, Carnegie Mellon University, hwiberg@andrew.cmu.edu, https://hwiberg.github.io/

## With Support From

Catherine Gill, Program Associate, CCC, Computing Research Association
cgill@cra.org

Haley Griffin, Senior Program Associate, CCC, Computing Research Association
hgriffin@cra.org

## Abstract

From 2021-2024, INFORMS, ACM SIGAI, and the Computing Community Consortium (CCC) hosted three workshops to explore synergies between Artificial Intelligence (AI) and Operations Research (OR) to improve decision-making. The workshops aimed to create a unified research vision for AI/OR collaboration, focusing on overcoming cultural differences and maximizing societal impact. The first two workshops addressed technological innovations, applications, and trustworthy AI development, while the final workshop highlighted specific areas for AI/OR integration. Participants discussed "Challenge Problems" and strategies for combining AI and OR techniques. This report outlines five key recommendations to enhance AI/OR collaboration: 1. funding opportunities, 2. joint education, 3. long-term research programs, 4. aligning conferences/journals, and 5. benchmark creation.

## Suggested Citation

Kulkarni, R., Brero, G., Ding, Y., Gupta, S., Koenig, S., Krishnan, R., Serra, T., Vayanos, P., Wasserkrug, S., Wiberg, H. (2025). AI/OR Workshop Series Final Report. Washington, D.C.: Computing Research Association (CRA).

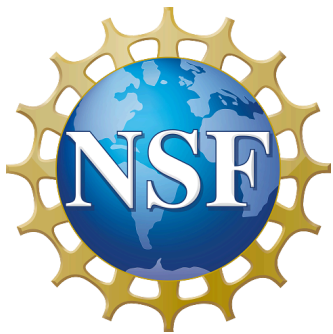

This material is based upon work supported by the National Science Foundation under Award No. 1734706 and 2300842. This cooperative agreement supports the Computing Community Consortium (CCC), a programmatic committee of the Computing Research Association (CRA).

Any opinions, findings, and conclusions or recommendations expressed in this material are those of the author(s) and do not necessarily reflect the views of the National Science Foundation.

### About the Computing Community Consortium (CCC)

A programmatic committee of the Computing Research Association (CRA), CCC enables the pursuit of innovative, high-impact computing research that aligns with pressing national and global challenges. Of, by, and for the computing research community, CCC is a responsive, respected, and visionary organization that brings together thought leaders from industry, academia, and government to articulate and advance compelling research visions and communicate them to stakeholders, policymakers, the public, and the broad computing research community.

# TABLE OF CONTENTS

# EXECUTIVE SUMMARY

Between 2021 and 2024, INFORMS and ACM SIGAI collaborated with the Computing Community Consortium (CCC) to host a series of three workshops, with a goal to explore ways to exploit the synergies of the Artificial Intelligence (AI) and Operations Research (OR) communities to transform decision making. These workshops aimed to create a unified strategic research vision for AI/OR that will maximize their societal impact in a world undergoing rapid technological transformation. By harnessing the complementary strengths of AI and OR, the workshops focused on overcoming the cultural differences between the two fields to pave the way for long-term collaborations.

The first two workshops of the series focused on recent technological innovations that enabled AI/OR collaboration, applications for AI and OR collaboration, methods for collaboration that ensure the development of trustworthy AI, and developing recommendations to enable these collaborations. The final workshop of this series concentrated on specific application areas that would benefit from integrating methods and techniques embedded in the disciplines of AI and OR, and how to enable these collaborations.

## Workshop Activities

The third workshop included presentations of "challenge problems," which require the collaboration of AI and OR researchers to efficiently and effectively solve them. These challenge problems were gathered through an open call for proposals to the community. Of the challenge problems which were proposed, 13 were accepted, which were combined into 6 topic areas. Breakout groups discussed the challenge problems and strategies for integrating AI and OR techniques.

- AI/OR Collaboration to Address the Opioid Epidemic and Causal Inference,
- Large Language Models (LLMs) and OR,
- Multi-Agent Interactions, Learning, and Risk Mitigation in AI-based Supply Chains,
- OR for Addressing Data Issues,
- Integrating OR and AI Through the Optimization Lens,
- AI-OR Trainers for Trustworthy Training of Non-Expert Humans/AI Agents, and
- The Tradeoff between Optimality and Explainability.

## Recommendations

This report outlines five recommendations for integrating AI and OR techniques and increasing collaboration between the two disciplines:

1. **Increased support for joint AI and OR research**, including new research institutes and funding opportunities that require co-PIs from both AI and OR backgrounds, to foster collaboration and progress in the field.

2. **Joint education opportunities**, such as summer schools, speaker series, joint degree programs and co-advising of PhD students.

3. **Long-term research programs** to accelerate progress in AI and OR collaboration by allowing researchers to work together in the same space for longer periods of time and focus on specific challenge problems.

4. **Aligning target conferences and journals** to reduce barriers to collaboration; Activities to further this effort could include establishing additional joint research conferences, educating researchers from both disciplines about high-quality publishing venues, aligning promotion and tenure policies, and clarifying policies on how OR journals will evaluate published conference papers at top ML conferences.

5. **Establishing joint benchmark datasets** that combine AI and OR and reflect complex human and societal dynamics to enable the comparison of algorithms and promote joint research between the AI and OR communities.

## 1. BACKGROUND

The Artificial Intelligence/Operations Research workshop series was born from a proposal (Das et al., 2020) to the Computing Community Consortium in early 2020 highlighting the importance of collaboration between the Operations Research (OR) and Artificial Intelligence (AI) research communities. This section outlines the motivation behind the proposal and articulates challenges and opportunities with such a collaboration. It is important to note that both the challenges and opportunities have intensified in light of the rapid advancements in Generative AI (GenAI) and other technological breakthroughs since 2020.

AI and Machine Learning (ML) have gained significant attention due to breakthroughs in game playing, computer vision, and natural language processing, surpassing human performance on specific tasks. While AI has a significant focus on ML, computer vision, and natural language processing, it also features research on constraint satisfaction, optimization, robotics, decision-making under uncertainty, human-machine collaboration, and multi-agent systems. The convergence of big data, high-performance computing, and AI has the potential to transform various sectors and address societal challenges. AI will continue to broaden its application domains and interact with other disciplines. The AI/OR workshop series organizers, understanding this inexorable progress, convened this workshop series to explore how the OR community can engage and collaborate with researchers in the rapidly changing field of AI.

The workshop series was the result of a collaboration between CCC, ACM SIGAI (ACM Special Interest Group on Artificial Intelligence), and INFORMS (Institute for Operations Research and the Management Sciences). This series aimed to establish a joint strategic vision for the AI and OR communities to maximize their societal impact by working together on problems of mutual interest. The two communities have complementary strengths: AI has adopted and refined optimization algorithms from OR, and OR has benefited from scalability challenges in problems arising from typical AI domains leading to reduced optimization solver times in OR. Decision-making under uncertainty is an active research area for both communities, including research in reinforcement learning, Markov decision processes and their generalizations, and stochastic optimization.

However, AI presents new challenges for OR, such as massive datasets, real-time autonomous operation, multimodal applications, and the need for robustness and reliability. These challenges present opportunities for OR researchers to develop novel optimization algorithms, methodologies for real-time problems and accounting for uncertainty, and integration of ML and optimization. While there are barriers and difficulties due to cultural differences, such as different vocabularies, missions, application areas, and investment imbalances, the workshops intended to explore opportunities to bridge these differences and build a strategic vision to address societal challenges by leveraging the strengths of both communities.

The workshop series aimed to overcome the difficulties and provide a stepping stone for strong and sustained collaboration between AI and OR along multiple aspects. It emphasized the need for a combination of data-driven (in AI) and model-driven (in OR) methodologies, pushing the boundaries of both research areas into uncharted and exciting territories. The OR community's expertise in developing mathematical models for specific applications and modeling languages can be combined with ML's ability to discover underlying structures from massive datasets. Additionally, the integration of ML and optimization, optimizing over multiple ML models, working with limited data, simulating potential future scenarios, and optimizing decision-making under uncertainty and risk are critical areas for synergy between the two approaches.

Importantly, AI and OR technologies need to promote equity, fairness, and absence of bias in their predictive and prescriptive models and algorithms. Fairness, equity, and absence of bias should be first-class citizens in the two fields, from data collection to ML algorithms, decision-making tasks, and validation.

CCC supported a series of three workshops to establish a joint strategic vision for AI/OR that will maximize the societal impact of AI and OR in a complex world with many challenging problems. Some of the following key goals were articulated:

1. To review the state of AI/OR along several axes.

2. To articulate several high-level research and education opportunities for each of these axes.

3. To identify potential grand challenges of mutual interest that would bring research and industry together from the two communities.

4. To select a few topics for summer schools and research programs to foster AI/OR collaborations.

Note that the above goals which were articulated in 2020 are still very relevant. Moreover, with the rapid developments in GenAI, we believe it is a particularly opportune time to highlight the imperative need for the two communities of AI and OR to collaborate towards solving deep and complex societal problems.

The three workshops were conducted between 2021 and 2024, supported by the CCC, INFORMS and the ACM SIGAI. This report provides brief summaries of the first two workshops, and highlights the real-world outcomes that emerged. Additionally, the later sections of the report detail the goals for the third workshop, practical applications that would benefit from AI and OR collaboration that were discussed during the third workshop, and findings and recommendations resulting from the entire workshop series.

## 2. WORKSHOP I: INTEGRATION OF OR AND AI METHODS AND APPLICATIONS

### 2.1. Summary

The first workshop in the series was held virtually (due to the pandemic) on September 23-24, 2021, and brought together a group of over 60 leading researchers in OR and AI to discuss their perspectives, opportunities and challenges from both Methods and Applications perspectives. Details of the sessions and resulting recommendations are available in the workshop report (Das et al., 2021). A few highlights are included in this section.

From a **Methods** perspective, leading researchers reviewed key technological innovations at the intersection of OR and AI. They presented the methodological overlaps and complementary strengths of AI and OR approaches, focusing on topics like constraint programming, optimization techniques, ML, and decision support systems. Their ideas explored how AI can enhance optimization algorithms to improve performance without sacrificing rigor and correctness, as well as how optimization techniques from OR can contribute to advancing the state of the art in AI. Some of the innovations in interface areas like multi-agent systems have benefited from developments in sequential decision-making and reinforcement learning which have co-evolved in the AI and OR communities, albeit with differences stemming from variations in language, approaches, and techniques.

Fairness, accountability, transparency, and ethics are crucial considerations in the application of OR and AI across various high-stakes societal decision-making domains. Researchers described how their work integrated techniques across a broad range of disciplines in AI, ML, OR, and theoretical computer science to advance the theory and practice of fair, transparent, and ethical algorithms used for decision-making. The theoretical innovations were driven by real-world problems in fairness-aware matching markets (e.g., for kidney exchanges), explainable planning and decision making, and quantifying uncertainty in ML models.

The **Applications** sessions explored various application domains where the integration of AI and OR techniques could yield significant benefits. Real-world use cases and challenges were described by leading researchers who worked with large organizations to solve tough commercial problems in areas such as supply chain resiliency, transportation, procurement optimization, price optimization, online resource allocation, etc. Several societal and policy-making applications were also shown to provide opportunities for collaboration in areas like public housing allocation, substance use prevention, biodiversity conservation, among others. The researchers who worked with communities and organizations to address important societal problems highlighted the fact that solutions must consider fairness, interpretability, and social impact. The problems in this category included HIV prevention, conservation, border security, addressing inequities in pay, etc.

The examples illustrated a few common themes:

- The need to blend data-driven approaches (common among AI applications) with a model-driven approach (as with OR applications).

- The importance of accounting for end-user needs in adopting these solutions.

- The challenges and opportunities in working with real data, whether they be offline and/or online or observational in nature.

- There is great power in blending AI and OR techniques to design interpretable and fair solutions that are trustworthy and mitigate existing biases.

Breakout discussions among the participants emphasized that while there are examples of collaborative research between the two communities, there is still ample opportunity to foster further collaboration and enable future researchers to overcome some of the inherent barriers that inhibit inter-disciplinary research, including challenges around data sharing and benchmark datasets, and human-in-the-loop decision making.

## 2.2. Key Recommendations

Below, we summarize the suggestions, concerns, and recommendations that arose from the first workshop:

- Key obstacles to collaboration among OR and AI communities include:

    o Differences in the publications culture between the two communities and lack of incentives to overcome these differences. Possible solutions include establishing processes for cross-publication in OR and computer science (CS) journals and conference proceedings; more specifically, IEEE journals might work as a venue to publish work from both AI and OR communities.

    o The lack of common language for similar techniques and methods shared by the OR and AI communities. This barrier can be mitigated by establishing competitions requiring the use of AI and OR techniques, summer schools, interdisciplinary courses in graduate programs, workshops and tutorials aimed at educating students and researchers from one community to the emerging research areas in the other.

- o   The differences in application domains and assumptions despite similar objectives, which have led to the development of different tools in OR and AI. However, the two communities have borrowed from each other to augment their respective sets of tools and methods. A question posed during the concluding session of the workshop was: *Given this adaptation and adoption of ideas from one discipline into another, the question is: "How best do we keep both areas abreast of such progress?"* Suggestions included attending each other's conferences and convening joint conferences between AI and OR researchers and practitioners. One such example of a successful conference worth emulating is CPAIOR (the international conference on the Integration of Constraint Programming, AI, and OR).

- Key recommendations for bringing AI and OR researchers together to realize the potential of collaboration at the data / methods / policy layers include:

    - o   Making data sets for a variety of applications available to both research communities and finding ways to overcome challenges with privacy that often exist with this requirement. Recent efforts in the AI community encouraging researchers to share data in the paper reviewing process (like AAAI - Association for the Advancement of Artificial Intelligence) may be a good model to emulate in other areas.

    - o   Fostering interdisciplinary collaborations through joint projects, events, and research groups of interest to both research domains. One such example at ICAPS (the top conference on AI planning and scheduling) was a train scheduling problem, which has traditionally been of great interest to OR practitioners. These competitions could be in application domains that can benefit most from AI/OR integration. Other suggestions included: Starting an ACM/INFORMS sponsored AI/OR conference or journal, establishing collaborative platforms to share ideas, and disseminating research best practices.

    - o   Starting an AI/OR summer school for AI and OR Ph.D. students, featuring both AI and OR expert lectures.

    - o   Exploring the creation of AI/OR MS and PhD programs with an aim of developing integrated hybrid methodologies combining AI and OR techniques.

    - o   Underscoring the benefit of including AI and OR experts in some policy discussions. Examples include supply chain risk mitigation, multi-modal transport, and mitigating public health risks/epidemiology. An important consideration in this arena is the fact that we do not always know the counterfactual. How can one address this issue?

    - o   Training students on ethics in AI because fairness and ethics play an important role in policy decisions as well as many applications in practice. Integrating ethics in core classes and teaching basic fairness metrics even in undergrad classes on ML could help instill the importance of these considerations in future researchers.

# 3. WORKSHOP II: AI/OR COLLABORATIONS FOR TRUSTWORTHY AI

## 3.1. Summary

The second workshop in the series was held on August 16-18, 2022, in Atlanta, GA on the Georgia Tech campus, with the support of the CCC, INFORMS, and ACM SIGAI. As with the first workshop, the second one focused on exploiting the synergies of the AI and OR communities and envisioning brainstorming activities to engage both communities and solve outstanding societal problems. As before, the speakers and participants were selected from AI and OR communities to foster a healthy exchange of ideas between the two groups. Specifically, the organizers aimed to identify promising research directions, discuss integration challenges and solutions, foster cross-pollination of ideas, and encourage joint problem-solving efforts.

A key decision from the first workshop was that we needed further discussion on Fairness and Ethics in AI and that we needed to consider the role of Causality in AI applications and explore the implications of human computer interactions while implementing automated solutions. Thus, the theme of the second workshop was Trustworthy AI with sessions related to Fairness, Human Alignment/Explainability/Human-Computer Interaction, Robustness/Privacy, and Causality/Explainable AI. The workshop was designed to study the state of the art, articulate grand challenges, and select topics for fostering AI/OR collaborations.

Details of the workshop sessions are included in the workshop report (Dickerson et al., 2023).

## 3.2. Key Topics Discussed During Workshop II

### *AI Initiative at the Federal Government Level*

Federal governments in the US and other countries have invested significantly in advancing AI research and created funding initiatives and programs to promote greater collaboration to solve consequential social problems. Some of the investment examples in the US include the establishment of the NSF AI Institutes (National Science Foundation, n.d.) and National AI Initiative Office (The White House, 2021), the launch of the National AI Research Resource Task Force (The White House, 2021), and the establishment of the National AI Advisory Committee (National Artificial Intelligence Advisory Committee [NAIAC], n.d.) among other AI related initiatives. As part of this effort all U.S. federal departments have been tasked with developing programs to advance the agencies' efforts on AI to harness the benefits and mitigate the risks of AI.

As an example of the efforts in various federal agencies, the workshop included a presentation from the Department of Transportation highlighting some of the key challenges to AI adoption and implementation, which are not unique to the DOT. These include explainability, liability, ethics and equity and other challenges which could be mitigated by maintaining a human-in-the-loop approach and promoting interdisciplinary collaboration across all disciplines. One of the key goals of the DoT's efforts is to play a leading role in data and AI strategies. Many of these activities have the potential to provide interesting problems for innovation at the intersection of AI and OR tools and techniques.

### Panel on Fairness

This panel broadly covered fairness in allocation, learning, and decision-making across healthcare, academic talent sourcing, government resource allocation, and political redistricting to combat gerrymandering. Discussions included the efficacy of using fairness as a constraint, new topics in allocation and market design, and fairness considerations in two-sided matching platforms like ridesharing.

Key takeaways included the need to consider fairness holistically across pipelines, the potential for unintended individual harms despite group fairness constraints, and the value of combining techniques from OR and ML to tackle complex fairness challenges.

### Panel on Human Alignment / HCXAI / HCI

This session covered various aspects of accounting for human factors when deploying AI/OR models for decision-making, including interpretability, explainability, model adaptability, and deviation from the status quo. Topics included aligning loss functions with decision objectives, understanding stakeholder preferences, measuring algorithmic bias, and designing effective explanations to calibrate trust in AI systems. The talks underscored the importance of thoughtfully accounting for human factors throughout the AI/OR pipeline.

### Panel on Robustness / Privacy

The panel covered several topics related to improving the trustworthiness, robustness, and privacy of ML and AI systems. Presentations covered trustworthy ML, robustness, privacy, and their interconnections across applications like healthcare insurance and the 2020 U.S. Census. Discussions revolved around the challenges of explainability, liability, model drift, privacy, security, ethics, and equity in AI adoption and implementation.

The panel highlighted the critical importance of robustness and privacy considerations when deploying AI/ML systems, and the need for thoughtful tradeoffs between privacy, accuracy/utility, and public trust.

Key takeaways from this session included:

- Improving robustness, generalization, and privacy can increase public trust in ML systems.

- Problem-driven approaches combining AI with areas like optimization can enhance robustness and fairness.

- Formal privacy methods like differential privacy are crucial for preventing disclosure of individual data, though enhancing privacy can degrade accuracy.

*Panel on XAI / Causality*

This session focused on explainable AI (XAI) and causality from computer science, OR, and engineering perspectives. Understanding causality provides valuable insights that enhance the explainability of AI methods. Talks covered adapting predictors under distribution shifts, design and analysis of panel data experiments, and causal inference applications in engineering contexts like wind turbine upgrades.

The key themes in this session highlighted the importance of causality in explaining AI models, including:

- Characterizing different types of distribution shifts from a causal perspective.
- Developing methods to estimate causal effects robustly under shift and interference.
- Applying causal inference to provide explanations and understand real-world impacts.

## 3.3. Key Recommendations from Workshop II

Many of the recommendations in this workshop echoed similar ones from the first workshop. These recommendations are enumerated below, in brief:

- Foster Interdisciplinary Collaboration
    - Encourage joint research projects, workshops, and conferences bridging AI and OR.
    - Develop interdisciplinary curricula and training programs.
    - Create competitions focused on developing efficient, equitable, and trustworthy algorithms; for example, related to ride-sharing services to address fairness disparities.
    - Establish an annual summer program providing multidisciplinary training in AI and OR to Ph.D. students, with joint lectures and a challenge problem requiring integration of techniques from both fields.
- Leverage Complementary Strengths
    - Combine ML, optimization, and decision-making methodologies.
    - Explore hybrid approaches capitalizing on AI and OR techniques.
- Address Ethical Considerations
    - Prioritize fairness, accountability, and transparency in system development and deployment.
    - Establish ethical guidelines and best practices.
- Facilitate Knowledge Sharing

- Create platforms for sharing data, code, and case studies.
- Foster open-source collaborations and reproducibility.

- Secure Funding and Support
    - Advocate for increased funding for interdisciplinary AI-OR research projects.
    - Engage with industry and government stakeholders.

- Promote Diversity and Inclusion
    - Actively promote diversity and inclusion within the AI-OR community.
    - Provide mentorship and support for underrepresented groups.

The workshop highlighted that through embracing interdisciplinary collaboration, leveraging complementary strengths, addressing ethical concerns, facilitating knowledge sharing, securing funding and support, and promoting diversity and inclusion, the AI and OR communities can jointly tackle complex challenges and drive impactful advancements across various domains.

# 4. OVERVIEW OF OUTCOMES FROM THE FIRST TWO WORKSHOPS

The discussions and resulting recommendations from the first two workshops yielded two specific outcomes:

- A Summer school described briefly in [Section 4.1](#).
- Outreach to other organizations described in [Section 4.2](#).

## 4.1. Summer School Held in 2024

The inaugural AI-SCORE summer school, held in College Park, Maryland, May 27 - June 1, 2024, was a significant event that brought together the OR and AI communities to explore the synergies between their respective fields. The first day of the program featured keynote lectures from eminent researchers, including Professors Michael Fu, Kevin Leyton-Brown, Tuomas Sandholm, and David Shmoys, who provided insightful overviews of the foundational concepts in OR and AI, and highlighted the progress made in bridging these fields. The ensuing panel discussion, moderated by the summer school chairs Lavanya Marla and Ferdinando Fioretto, involved a vibrant discussion with students on the future of the intersection of these areas, as well as practical and career aspects of blending the OR and AI communities. AI-SCORE (n.d.) provides more details on the schedule.

The following days of the summer school offered modules on Reinforcement Learning and Fairness, led by faculty from both the OR and AI areas. Reinforcement Learning was led by Profs. Vivek Farias and Scott Sanner (days 2 and 3), and Fairness was led by Profs. Siddhartha Banerjee and Aaron Roth (days 4 and 5). Professor Sven Koenig discussed reinforcement learning in the context of the upcoming League of Robot Runners competition and inspired students to build on their knowledge from the summer school towards the competition. Both modules were well-received and fostered active participation from students working in cross-disciplinary teams. This setup effectively promoted the exchange of ideas and the development of new methodologies that leverage both model-driven and data-driven approaches. Overall, the goals of the AI-SCORE summer school were well-achieved, as evidenced by the high level of engagement and solutions proposed by the participants. Student feedback was strongly positive, emphasizing the need for further iterations of the school and the need to bring OR and AI towards a common platform. The school aimed to not only train the next generation of researchers to be conversant in both OR and AI methodologies but also posed exciting challenge problems that spurred significant interest.

Looking ahead, future editions of the AI-SCORE summer school are planned to continue fostering collaboration between both fields, with a focus on further integrating computational paradigms and algorithmic approaches to improve interoperability, trustworthiness, and privacy in decision-making.

## 4.2. Outreach to Other Professional Organizations

To further enhance the collaboration between OR/MS and AI, a Memorandum of Understanding (MOU) has been signed between INFORMS — the largest professional OR/MS organization, and AAAI - the largest professional AI organization. The goals and objectives of this MOU include:

1. Driving forward decision making through the combination of AI and OR/MS skills and capabilities.
2. Showcasing these capabilities in important application areas to be agreed upon jointly between AAAI and INFORMS.
3. Enabling members of AAAI and INFORMS to be aware of and apply the appropriate combination of AI and OR/MS skills relevant to our shared goals.
4. Fostering responsible application of AI/OR/MS techniques.

An exciting result of this collaboration with AAAI is the [Bridge program](), titled, "Combining AI and OR/MS for Better Trustworthy Decision Making", at the AAAI conference in February 2025.

Additionally, an INFORMS wide ad-hoc committee on AI was formed to actively explore collaborations between INFORMS and additional AI organizations, and what can be learned by collaborating with other professional organizations such as the Society for Industrial and Applied Mathematics (SIAM), the American Mathematical Society (AMS), and the Institute of Mathematical Statistics (IMS) who have been impacted by AI.

# 5. WORKSHOP III: GRAND CHALLENGES FOR AI/OR COLLABORATIONS

## 5.1. Ideas Resulting From the First Two Workshops Leading to Workshop III

In the first two workshops, we explored how cross-pollination of OR and AI methods and techniques could benefit each discipline, and approaches to enable this collaborative understanding. The next question is, "What are the compelling 'big' ideas that require AI and OR collaboration?" What are the newer challenges which provide opportunities for innovation at the intersection of the two communities?

The recent advent of Generative AI prompts new use cases for OR and AI collaboration as well. How can researchers combine modeling, planning and AI generative capabilities for the early problem formulation phases, through to the implementation and deployment of solutions? Some related questions are:

- How can researchers certify AI systems to be reliable and ready for deployment?
- How can researchers audit and certify the provenance of data used in AI?
- How do we accomplish this in the context of societally consequential applications?
- What is needed for the theory and empirics around operationalizing AI?

## 5.2. Goals for Workshop III

In the third workshop, the organizers aspired to solidify the learnings from the first two workshops, namely, the state of AI/OR along several axes and the high level research and education opportunities for each of these axes.

With the rapid developments in GenAI, it was a particularly opportune time to convene the third workshop to delve into a few specific grand challenges and produce a final report highlighting the imperative need for the two communities of OR and AI to collaborate towards solving deep and complex societal problems. We aimed to highlight the different strengths and perspectives that the two communities bring towards solving the problems, enabling greater progress than working in their own silos. This goal is a key reason why such a workshop was so timely and of great importance to advance the research directions in AI.

We also hoped to engage NSF program directors by emphasizing the enormous potential in combining the researchers from OR and AI to advance the use of AI to solve complex, consequential decision problems in society. The discussions in the first two workshops highlighted the interest in collaboration between these two communities and the possible opportunities by bringing them together with different perspectives and modeling approaches to solve challenging problems and provide guidance for the future development of AI systems. Finally, based on these earlier discussions and some of the changing landscape since the start of the workshop series, we hoped to provide a list of recommendations for the future.

We planned the third workshop around a set of grand challenge problems solicited from the OR and AI communities, seeking input on the following questions:

- What are challenges that you believe must be faced to tackle this problem, and what are the mechanisms needed to address them?

- How can policies/mechanisms incentivize collaboration between the two communities?

- Why do you think the two communities need to collaborate to address this challenge? For example, what AI tools (e.g., ML, reinforcement learning, LLMs, GenAI) and what OR tools (e.g., robust optimization, queuing theory, integer optimization) are needed?

- How do you create systems and models that assure trust in AI? Can we combine the classic planning and modeling approaches with generative approaches in a meaningful way?

- Can we articulate a summary of findings and recommendations that is usable by government agencies and science and technology policy makers for setting priorities?

- Can joint research provide a plan for mitigating the risk of AI systems?

- Do we need educational workshops / summer schools to foster such combined research activities?

## 5.3. Challenge Problems for AI/OR

*1. Causal Inference and the Opioid Epidemic*

The opioid crisis in the United States has reached alarming proportions, as evidenced by staggering statistics. In 2021, the number of individuals who lost their lives due to drug overdoses surpassed six times the figures recorded in 1999. Even more concerning, the year 2021 witnessed a distressing 16% surge in drug overdose (OD) deaths compared to the previous year, with opioids implicated in over 75% of the nearly 107,000 fatalities. A deeper analysis of the crisis reveals a complex landscape: while opioid-related death rates increased by over 15%, prescription opioid fatalities remained constant. On a more hopeful note, heroin-related deaths decreased by nearly 32%, but synthetic opioids (excluding methadone) saw a troubling rise of more than 22%. This crisis has become a critical issue demanding urgent attention and comprehensive solutions (Centers for Disease Control and Prevention, 2023). In Massachusetts, the death rate increased to 33.5 per 100,000 people in 2022, 2.5% higher than in 2021 and 9.1% higher than the pre-pandemic peak in 2016.

Addressing the opioid crisis often necessitates working with observational datasets often failing to satisfy classical causal inference assumptions made in the causal inference literature. By blending OR and AI, one could help address the limitations of classical causal inference and enable the use of this observational data to inform better decision-making in several domains critical to this crisis. To ensure robustness, fairness, equity, and address data challenges (e.g., dimensionality, sparsity, heterogeneity, missing data, etc.), incorporating domain knowledge, trust, interpretability as well as computing scalability, the integration of AI and OR is must. This integration will enable public and private sector policymakers and decision-makers to make unbiased and reliable decisions and laws by uncovering the secrets of big data, potentially saving millions of lives and scarce resources. The integration of AI and OR has already helped mitigate uncertainty in causal inference (Morucci et.al, 2022), the curse of dimensionality (Islam et.al, 2022), bias and model dependency (Islam et.al, 2019), and computational complexity induced by big data.

Recent progress has indicated that causal discovery is surprisingly feasible, even in scenarios where the given data are Missing Not at Random (MNAR). This is surprising given that structure learning (even non-causal) was almost exclusively limited to the relatively simplistic assumption that data is Missing at Random (MAR). This progress suggests that the scope of causal discovery may in fact be extended to real-world scenarios that were up to this point considered out of reach.

The ability to achieve these goals depends in part on advances in the mathematical formalization of these scenarios, including their representation as Integer Linear Programs (ILPs), but possibly going beyond linear constraints as well. Subsequently, computational advances are required to facilitate the discovery of the solutions to the resulting programs, which includes the adaptation and integration of existing AI and causal discovery algorithms with tools and techniques from OR and ILP.

Additionally, data issues, especially assumptions made in the OR and AI literature about data quality, continue to burden researchers using data and algorithms to help address the opioid epidemic. In particular, making inferences about counterfactuals or discovering causal relations in the context of this high stakes domain (which is needed to make predictions and prescriptions for the future) requires large data and high quality data. Unfortunately, the datasets currently available to help make predictions and informed decisions and policies in this context are permeated by data issues, including the small amount of data available, the low quality of the data, which often has missing entries, selection bias, and even more critically, presents unobserved confounders (covariates that were used historically to select individuals into treatment and that are not recorded in the data, limiting our ability to eliminate selection bias using standard methods). These limitations of the data prevent the use of standard techniques which, if applied, may result in decisions and policies performing poorly.

Below, we identify several areas of research and projects which, with dedicated researchers and funding, would be able to significantly improve both drug overdose and addiction rates in the United States.

**Decision Analytics Tool for Personalized Prescription**

Predictive models, which leverage large-scale healthcare data with AI to proactively identify individuals who are at a heightened risk of developing Opioid Use Disorder (OUD) (Hasan et al., 2021), are needed to develop care methods which prevent addiction before it occurs. This predictive outcome could be very useful in identifying the right treatment options (e.g., medication and dosage), given that each medication comes with its own risks and benefits. We propose to use a multi-objective decision analytics framework, integrating AI, OR, and knowledge from domain experts. This tool could serve as a vital resource for healthcare practitioners, aiding them in generating personalized medications based on each patient's unique profile and risk of OUD.

**Prescription Opioids Diversion Reduction Policy**

A fundamental problem with the opioid epidemic crisis is that many patients get prescribed too many opioids and don't use up every pill. Additionally, they don't dispose of the surplus drugs properly, potentially facilitating the diversion of prescription opioids to secondary users and the black market. AI could be used to understand how the OUD/OD crisis varies across different geographical locations, communities, and socioeconomic demographics. It can also help identify and rank the critical regions. Subsequently, we can develop an optimization framework to identify optimal policies for allocating resources, such as installing disposal kiosks and providing motivating education/training and incentives (Hasan et al., 2023).

**Design Optimal Allocation Policies of Overdose Prevention Toolkit**

To mitigate the risk of fatal and non-fatal ODs, the presence of a prevention toolkit is a lifesaving measure. By integrating AI and OR, we can construct a decision analytics framework to allocate these lifesaving toolkits most effectively to the communities that require them the most. This process involves utilizing AI to analyze historical data, monitor real-time information sources, and create predictive models for identifying high-risk areas and individuals. Subsequently, we optimize the distribution of the toolkit, offer training, and consider factors such as risk assessment, response time, and resource availability.

**Design Policies for Improving Treatment Retention**

OUD is a treatable condition, and one of the FDA-approved medications for OUD is buprenorphine/naloxone. Successful treatment with buprenorphine/naloxone has been linked to reduced morbidity and mortality in OUD patients. Unfortunately, many OUD patients discontinue buprenorphine treatment prematurely. Medication adherence during buprenorphine treatment is often inconsistent among patients. We could use AI, leveraging claim data, to identify predictive factors such as demographic information, addiction severity, comorbidities, and treatment adherence patterns that contribute to treatment discontinuation. These models can then provide real-time risk assessments for individual patients, enabling healthcare providers to proactively intervene when the risk of discontinuation is high. A decision analytics framework, integrating reinforcement learning, could be developed to optimize resource allocation for better treatment retention and improved patient outcomes.

## 2. Bringing Together Generative AI and Operations Research/Management Science

Today's Generative AI (GenAI) models, including Large Language Models (LLMs) such as ChatGPT (OpenAI, n.d.) and Gemini (Gemini, n.d.), are the culmination of decades of AI research and are providing amazing, advanced AI capabilities through natural language interaction, or prompting. Millions of people are already using these models, and this usage is only expected to grow, potentially transforming many knowledge and expertise-based jobs and professions (Dell'Acqua et al., 2023), and, ultimately, our society. The proliferation of these models and the ubiquitous use of them present many challenges as well as opportunities, described in this section.

**Generative AI Issues Relevant to OR/MS – Challenges**

The pervasiveness and accessibility of GenAI results in new risks and issues relevant to OR/MS. These include:

- **Dangers of Using GenAI for decision making:** As its usage continues to grow, GenAI will have an increasing influence on decision making. However, while GenAI models exhibit some reasoning capabilities, at their core, they are probabilistic ML models. Therefore, if GenAI is being used, for example, to directly generate solutions to optimization problems such as in (Kool, van Hoof, & Welling, 2018), it is unclear how close to optimal the solution is because such networks are very difficult to analyze analytically. Moreover, GenAI currently has severe limitations in its ability to handle complex constraints when processing optimization problems. Therefore, ultimately it is unclear how good or bad the decisions based on GenAI will be. Moreover, users tend to be over-dependent on GenAI outputs (Dell'Acqua et al., 2023). These limitations, coupled with GenAI's *hallucinations* (erroneous, yet highly convincing outputs) and an absence of human oversight may further compound the dangers of incorrectly using GenAI in general, and as part of a decision-making process in particular.

- **Inherent issues and limitations of GenAI models:** The proliferation of GenAI models raises many issues and concerns. GenAI models are statistical machines typically trained to optimize some loss function using a combination of semi-supervised ML and reinforcement learning with crowdsourced human feedback. This training method necessarily limits supervision. Therefore, it is unclear how aligned these models are with individual, organizational and societal goals. For example, they can be used to generate highly convincing disinformation, used for security attacks, etc. Moreover, often systems are composed of multiple GenAI models, where each GenAI model has its own loss function. Such composition may further compound these issues. GenAI models also raise new intellectual property and privacy rights.

**OR/MS Issues Relevant to GenAI**

The goal of Operations Research and Management Science (OR/MS) is to enable better data driven decision making based on mathematical models grounded in strong theory. Typically, these models explicitly target optimizing well-defined measurable goals defined by an objective function. These techniques are well suited to handle cases in which there are a large number of possible decisions, and in which the decisions are subject to complex constraints.

While OR/MS has proven itself to be extremely valuable in many scenarios and use cases, its application requires significant expertise and skills. Thus addressing a real-world decision-making problem using GenAI models coupled with OR/MS techniques requires an OR/MS expert to work with an organizational decision maker to understand the problem and create appropriate formal models. These OR/MS experts need to work with organizational decision makers to validate the GenAI model and its appropriateness for real-world problems. Finally, as real-world problems often result in optimization problems which are NP-Hard, then a) the model created by the OR/MS expert needs to be an effective mathematical model (i.e., a model that is formulated in a way that can be solved by the current OR technology) and b) the parameters of the OR solver need to be set appropriately for the specific model.

**Combining GenAI and OR/MS**

We propose to combine GenAI and OR/MS to:

- Enable improved decision making through the combination of GenAI and OR/MS
- Address many core issues of GenAI, such as safety and alignment with human values, through the application or OR/MS techniques

The following subsections list some of the challenges and opportunities regarding these.

*Opportunities and Required Research in Combining GenAI and OR/MS*

**Combining GenAI and OR/MS for Better Decision Making**

In this section, we detail two directions for combining GenAI with OR/MS for better decision making that arose out of the third workshop:

- Democratizing optimization, i.e., enabling decision makers with little or no optimization expertise to utilize optimization to make better decisions

- Teaching better decision making

*Combining GenAI and OR/MS for Democratizing Optimization*

The ultimate democratization of OR/MS for decision making would be to enable business users interacting with GenAI in natural language to address a business decision making problem. The user will then be able to provide feedback on solutions, which will subsequently be used to improve the models. In effect, this capability will have to address all the tasks currently carried out by OR/MS experts, including the ability to translate a business level problem description into formal models, interacting with the business user to validate and refine the model, creating efficient models, and tuning the OR solvers. GenAI has already demonstrated capabilities relevant for such democratization, including the ability of LLMs to generate optimization models from precise natural language descriptions, such as those shown in (AhmadiTeshnizi, Gao, & Udell, 2023) and (Ramamonjison et al., 2023). The ability of LLMs to tune solver parameters (Li, Mellou, Zhang, Pathuri, & Menache, 2023) and using LLMs as a natural language interface to interact with and add constraints to existing optimization models (Lawless et al., 2023) are also methods used by GenAI for democratization. However, as is shown by (Wasserkrug et al., 2024), much work still remains in enabling full democratization. Enabling such democratization will require the appropriate creation, adaptation, and composition of GenAI models, together with OR/MS knowledge, models, and techniques.

*Combining GenAI and OR/MS for Teaching Better Decision Making*

Combining GenAI and OR/MS for educating people in data- and model-driven decision-making practices can be done in various ways. For example, LLMs augmented with or adapted to specialize in OR/MS knowledge could be prompted to assume the role of a decision-making tutor or mentor. An alternative approach could involve leveraging GenAI and OR/MS to develop a series of scenarios designed to train individuals in effective decision-making. In such an application, the GenAI models would generate a curriculum of scenarios, while the OR/MS models and techniques ensure that the scenarios are designed to promote valid and safe behaviors for trainees. Another approach would be to adapt GenAI models (for example, through prompting), to help human decision makers in critically analyzing the output of other GenAI models used for decision making, thereby educating individuals in critical thinking relevant to the decision making process.

*Using OR/MS to Address Core GenAI Issues*

In addressing some of the core issues arising from GenAI, OR/MS could:

> (a) Provide proposals for new policies related to the safe and trustworthy development and usage of GenAI,
>
> (b) Estimate or predict the effect of these policies, and
>
> (c) Provide methods to operationalize these policies.

For example, an important policy domain for LLMs is on the intellectual property (IP) of the data used to train these models. Using contract theory, mechanism design, and game theory, coupled with LLMs' knowledge, may help develop new data contracts among the users of LLMs, the data providers for these systems, the companies/entities that own these LLMs, and the policymakers. New OR/MS models may also help understand the economic effects of LLMs on certain labor markets. Finally, OR/MS models may help compose multiple GenAI models into an application aligned with an overall target goal.

*Data Requirements*

Adapting existing GenAI models or training new ones to include OR/MS methodologies will require both appropriate data and suitable metrics to measure progress. For example, democratizing optimization models from business problem level descriptions will require a set of optimization problems, each containing a correct and efficient "ground truth" optimization model. In addition, it will require some simulation of a business user. Here too, GenAI may be able to help, either by accelerating the creation of data by extracting it from articles containing use case descriptions and corresponding models (such as articles appearing in the INFORMS Journal on Applied Analytics (n.d.)) or using an LLM to play the role of a user attempting to solve a problem through optimization.

## 3. Understanding Multi-Agent Interaction and Learning in AI-Powered Supply Networks

The presence of algorithmic agents is rapidly increasing across markets. Supply chains offer a prime example of this trend. The 2020 pandemic significantly accelerated the adoption of algorithmic agents in these markets (Andrade, Frongillo, & Piliouras, 2021), highlighting the vulnerability of traditional supply chains to geopolitical disruptions, with entire organizations experiencing rippling effects (Carbonaro, 2022). In response, many companies are transitioning to digital solutions, particularly AI, to optimize their operations. Consequently, the global supply chain AI market is projected to reach $13.5 billion by 2026 (Javaid, 2022), and supply chain organizations anticipate a twofold increase in their level of machine automation by 2028 (ThroughPut, 2022).

Automated software agents that better forecast and adapt to changing business environments offer great potential. Major consulting firms are already offering related services, and business schools are providing new courses on the subject (The Economist, 2019). ML and more specifically reinforcement learning has drawn considerable attention as a paradigm for implementing learning software agents that adapt to changing business environments for tasks such as pricing, replenishment, order management, and supplier selection (Giannoccaro & Pontrandolfo, 2002; Rana & Oliveira, 2014; Rolf et al., 2022; Sutton & Barto, 2018). However, we do not yet understand the dynamics that emerge from self-interested and learning agents in such supply chains.

Many supply chain management (SCM) problems (de Kok & Graves, 2003) are characterized by the interaction of several decision-makers in partially-cooperative and partially-competitive settings. Therefore, game theory has been used to describe the outcome of strategic interactions, usually in the form of equilibrium predictions (Cachon & Netessine, 2006; Grauberger & Kimms, 2018). These results are based on strong informational assumptions about the participants and the environment, which frequently do not hold true in SCM contexts. Moreover, it has been shown that equilibrium computation (e.g., Nash) can be hard (Daskalakis, Goldberg, & Papadimitriou, 2009), and it is well known that there can be many equilibria even in the relatively simple setting of static normal-form games. Most real-world planning problems in supply chains (for example, transportation planning or inventory management) are dynamic or have multiple stages. Equilibria in such games are even harder to compute (Daskalakis, Golowich, & Zhang, 2022), and it is far from obvious that automated agents would find or play such an equilibrium.

Rather than rely on game-theoretical equilibrium strategies from the start, automated agents in supply chains can use ML to adapt to their environment. The input in multi-agent learning, however, is non-stationary because it depends on the strategic behavior and learning of other agents, which makes it very challenging to analyze. In spite of the challenge, it is essential that we work to better understand the interaction of multiple learning agents. In competitive pricing problems, for example, algorithmic collusion has caught the attention of regulators, because there is evidence that learning pricing agents might collude at high prices (Calvano, Calzolari, Denicolo, & Pastorello, 2020). Ultimately, we ask an age-old question in the economic sciences, dating back to Adam Smith: Will the interaction of self-interested autonomous learning agents lead to equilibrium? And if so, will the outcome be socially desirable?

To address these questions, it is crucial to incentivize research focused on better understanding the dynamics of learning agents within supply chain game models. Specifically, we aim to identify conditions under which agents converge to an equilibrium, cycle, or exhibit chaotic behavior. If learning agents converge to an equilibrium, then such an equilibrium may be a credible prediction of the outcome of a game. If not, perhaps the learning agents' behavior can be influenced to yield efficient equilibrium outcomes.

Another promising avenue is to investigate how to design supply chain markets where learning agents are expected to converge to desirable outcomes. This approach was demonstrated by (Johnson, Rhodes, & Wildenbeest, 2023) in the context of algorithmic collusion, where they showed that the collusive pricing behaviors identified by (Calvano, Calzolari, Denicolo, & Pastorello, 2020) can be mitigated by implementing market rules favoring lower-priced products. While their method for deriving these rules was analytical, (Brero, Mibuari, Lepore, & Parkes, 2022) later introduced an alternative set of anti-collusive rules that build on the ML techniques for automating rule derivation, initially introduced by (Brero et al., 2023). These techniques for developing favorable market rules could potentially be adapted for supply chain models to promote designs that ensure desirable outcomes among learning agents.

The understanding of automated supply chains is central to future economies and the welfare of nations. Effective supply chains are also crucial for the development of GenAI systems that rely on "foundation models" like GPT-3. These models act as the base layer in a multi-level AI supply chain, feeding into more specialized AI for specific applications. This creates a branching structure where the impact of base models amplifies through the chain. Developing and deploying these systems involves collaboration between various players throughout the chain, from data creation to user deployment (Bommasani et al., 2021).

In conclusion, the challenges and opportunities presented by the adoption of autonomous software agents in supply chain markets create a valuable opportunity for collaboration between the AI and OR communities to promote the responsible and fair development of these technologies (Black et al., 2023; Cobbe, Veale, & Singh, 2023; Papachristou & Rahimian, 2023).

## *4. Data Scarcity and Privacy in Healthcare*

Machine learning algorithms are increasingly prevalent in high and low-risk settings, providing predictions that can be used within existing decision-making frameworks. In a healthcare setting, ML can analyze vast patient data, allowing subsequent models to merge seamlessly into organizational workflows. For example, ML models are now used to anticipate unplanned hospital readmissions or to predict the length of stay in inpatient units, improving both the operational and clinical outcomes of the system. ML fundamentally relies on good data for both training and evaluation; however, in many settings, access to high-quality data is challenging. Several hurdles related to curating robust datasets have been highlighted by other challenge problem groups within this report, such as the issue of unobserved confounders in observational data. In this group, we highlight two additional data issues that the AI and OR communities are uniquely positioned to tackle collaboratively: data scarcity (in training and evaluation) and privacy. Although this group focused on healthcare, these techniques apply across industries.

Data scarcity is inevitable in impact-oriented ML applications. Many public sector or non-profit organizations are highly resource-constrained with limited data availability. Organizations often lack access to large retrospective datasets, which impacts both internal model development and the adoption of externally trained models. Internal development of models is hampered by scarce training data. While adapting a model which has been developed and effectively deployed at a source site for use at external sites might seem like a promising solution to this type of scarcity, it is often hindered by scarce validation data. In such settings researchers must continue to assess model performance and potentially make modifications to make it suitable for deployment at external sites. However, external sites often have limited data and resources. This makes it challenging to measure the degree of data drift between the original derivation population and the target population at the new site. Furthermore, most complex ML algorithms do not provide theoretical guarantees of performance as a function of the degree of data drift between the distribution of the data on which the algorithm was trained and subsequently validated.

Even when data are available, data privacy is a key concern, especially in sensitive settings like healthcare. Many models have been trained on data where the models have been shared or released to the public, but the training data are kept private. Privacy could be desirable for many reasons, such as if the data contains trade secrets, copyrighted information, or protected personal health records. As an example: in federated learning for health data among several hospitals, it is often assumed that the data within each institution can be kept hidden, but this is not necessarily true.

The highlighted data issues have significant implications on patients and, in particular, they hinder equitable access to ML innovation. Limited data are particularly relevant in low-resource settings, so current model development and deployment significantly favor well-funded, high-volume systems and disadvantage low-resource, smaller community health centers. A better approach to model application in low-data settings would democratize the important ML work being done in healthcare systems globally. It would allow for broader adoption, increasing the reach of high-performing models and the resultant improvements in care. Additionally, insufficient data privacy in model building also has significant downstream impacts. It can be detrimental to patients by inadvertently exposing their data, which in turn erodes trust in the development of ML models. This can make people reluctant to share their data, thus reducing sample sizes and potentially leading to the under-representation of certain groups in the training data. Both factors harm the quality and generalizability of the model.

This challenge area presents an opportunity for the OR and AI communities to collaborate, combining research in data augmentation, privacy-preserving models, model generalizability, and decision-focused learning frameworks. Both communities have existing literature that can be leveraged, yet these data issues remain unsolved: effective solutions will integrate both OR and AI. Below, we outline concrete areas that leverage both communities:

**Data Scarcity – Model Training**

Small training data can be addressed in multiple ways, such as data augmentation, exploiting side information, and task-based learning (i.e., learning with respect to decision loss). Embeddings are commonly used to represent data in a lower-dimension space that facilitates learning. Fine-tuning pre-trained embeddings to new data, or "robust embedding", allows researchers to efficiently learn in small data regimes. Critically, most predictive tools that are developed for healthcare systems aim to improve the decision-making process for a downstream prescriptive task, which is a key focus of the OR community (Mišić & Perakis, 2020). Decision-focused learning (Elmachtoub & Grigas, 2022; Ferber et al., 2020) in a sense simplifies supervised learning tasks by only penalizing errors that change a decision, rather than all prediction errors that do not affect decisions. This can be advantageous in limited data settings.

**Data Scarcity – External Validation**

The AI community has made significant progress in quantifying model performance for unseen data, both empirically, leveraging Generative Adversarial Networks (GANs) (Bertsimas & Orfanoudaki, 2021; Creswell et al., 2018), and with theoretical guarantees, through conformal prediction (CP) (Angelopoulos & Bates, 2021; Barber et al., 2023). Beyond measuring model performance, both communities have made progress in improving models for application in new settings, often with limited data. Examples include transfer learning (Mustafa et al., 2020), federated learning (Sarma et al., 2021), and mixed-integer optimization-based stable learners (Bertsimas & Paskov, 2020).

**Data Privacy and Protection**

Data protection requires understanding the vulnerabilities of models and systems to side attacks, to devise better defense strategies. This requires expertise in both ML models and optimization algorithms and theory. The optimality landscape of ML models is still only partially understood. Some theory exists for certain limiting special cases such as very shallow models with few hidden layers, or "infinitely" wide hidden layers based on the neural tangent kernel. In real life, most networks are neither shallow nor infinitely wide, and the training data are not independent and identically distributed, so there are theoretical gaps that must be addressed.

## 5. Integrating OR and AI Through the Optimization Lens and the Tradeoff Between Optimality and Explainability

One of the issues fundamental to all application problems discussed in the earlier subsections of this report is the effective and capable optimization solution methods that underpin all AI methods. Combining AI and OR can improve upon "traditional" methods for addressing complex problems. ML and AI have made great strides in applications where there is a large amount of data and the key performance metric is computationally efficient to evaluate, even when the feature space is large. Much of this progress has been backed by the use of optimization algorithms in training ML models. As a result, improvements in optimization algorithms and OR techniques in general would directly benefit further advances in ML and AI. Conversely, AI has been successfully employed to improve the performance of optimization algorithms.

Optimization problems that are either discrete or highly nonlinear, non-smooth, and non-convex can be challenging to solve, even approximately. With further uses of optimization emerging in AI and ML, such as in the context of formal verification and counterfactual explanations, better optimization algorithms are also fundamental for the progress of AI. In specific applications of optimization, in which similar versions of a problem are solved from time to time, AI that is capable of designing heuristics customized to each application's own objective function, constraints, and data, could exploit the structure in the historical data to bring unprecedented speed-ups to optimization. OR can help bring in tools such as Bayesian optimization and help exploit the distributional information available with each iteration, as well as exploit structural properties from previous solves.

In addition, there are hybrid combinations of AI and OR which defy their conventional boundaries. One such example is neural surrogate models, or simply constraint learning, in which either an optimization model is embedded within a neural network or a neural network is embedded within an optimization model. The potential for this integration has been illustrated with applications on maximizing scholarship acceptance based on student profile; toxicity mitigation while optimizing patient survival in chemotherapy treatments; optimizing power generation and voltage regulation in power grids; and the control of automated operations such as room temperature regulation.

Moreover, both AI and OR may benefit from further synergy with other disciplines, such as physics. For example, energy models have been previously used to explain classification and optimization problems separately. Another example is the use of language models to solve differential equations symbolically, which may suggest the potential of generative models for tackling optimization problems in the future. Along with these advances, a key challenge is the explainability of such models in high-stakes settings, where guarantees from an optimization framework may not be easily accessible when AI is added.

Bringing the AI and OR communities together helps each contribute to the other more frequently, such as through more nuanced application of ML to optimization algorithms, as well as more efficient optimization algorithms for ML needs. Constrained optimization supports logical and physical constraints in ML applications. Exact algorithms in optimization may also provide explainability for ML models obtained with them. High-risk applications may need a more nuanced application of optimization in training by weighting false positives and false negatives differently.

## 6. RECOMMENDATIONS FOR FUTURE ACTIONS

Ultimately, this workshop series sought to strengthen collaboration between the AI and OR communities. While this report specifically focuses on the AI and OR communities, our recommendations more broadly apply to collaboration with related disciplines, including causal inference, economics, statistics, and others. In all cases, collaborations rely on having funding, data, and interesting problems to work on. They require further effort when bridging two communities: interaction (i.e., opportunities to meet and work together) and incentive alignment. Several recommendations emerged from the workshop, which span across the challenge problems. The recommendations range from local university efforts to national funding opportunities.

## 6.1. Funding Opportunities

We recommend establishing new institutes (or influencing the agenda of existing institutes) for joint research along the lines described in Section 7. Additionally, we propose the creation of funding opportunities that require the participation of co-PIs from both AI and OR backgrounds, incentivizing new collaborations that will contribute to fundamental progress in the field.

Progress depends on the combined expertise of researchers in optimization theory and algorithms, scalable numerical algorithms, and scalable ML technologies. Ideally, funded research centers would allow collaboration between AI and computational optimization specialists in an environment with massive computing capacity and access to data. Joint research opportunities that will benefit from increased collaboration include the topics and real world challenge problems in Section 7.

Currently, funding opportunities remain a challenge as proposals are still evaluated either by panels on AI or OR/MS. Funding programs geared towards this interface, and correspondingly jointly evaluated, could give this research a significant boost. A long-term funding commitment will increase research in the area and help make systemic change to grow research at the interface of the two disciplines.

## 6.2. Joint Education Opportunities

Successful collaboration between the AI and OR communities requires a shared understanding of our respective fields. Each group can benefit from learning about the unique benefits and values afforded by the models and techniques of the other group. We propose to create an annual summer school that trains PhD students on topics at the intersection of these fields. Indeed, most OR students have less exposure to AI methods and working with large-scale data. Similarly, most CS students are not trained in the OR methods needed to address the policy and decision-making questions that arise in this domain (e.g., ILP, robust optimization). The AI-SCORE summer school, described in Section 4, provides a prototype for what a recurring program might look like going forward.

We also propose ongoing opportunities for interaction and shared education throughout the academic year. We propose the creation of a regular speaker series (along with recurring workshops at AI conferences such as AAAI, IJCAI (International Joint Conference on Artificial Intelligence) and NeurIPS and at the INFORMS Analytics Conference and Annual Meeting) for researchers at the intersection of the two fields. We also recommend creating opportunities for professional networking to promote interdisciplinary collaboration.

Beyond society-level national efforts, universities and academic departments can also lead local initiatives. Joint degree programs within universities can provide a formal path for interdisciplinary training. One example of interdisciplinary graduate programs is the [interdisciplinary certificate program launched by the University of Southern California](#) for PhD students joining either the Computer Science or the Industrial and Systems Engineering Department at USC. We recommend normalizing and facilitating co-advising of PhD students between faculty that are experts in OR and experts in AI. This can play a very important role to create a generation of experts at the interface of the two fields and, through joint publications, help showcase the power of integrating OR and CI to address important societal challenges.

## 6.3. Long-Term Research Programs

AI and OR researchers generally sit in different departments and have limited formal cross-institution interaction due to their separate major conferences. We believe that dedicated collaborative research workshops could accelerate progress. We propose the organization of long-term research programs, like the one recently organized at the [The Institute for Mathematical and Statistical Innovation](#) (IMSI) at the University of Chicago, allowing researchers from diverse backgrounds and fields to work together in the same physical space over more extended periods of time. This research program could contain tracks for selected challenge problems, with the goal of actively working on research topics that can culminate in publications and/or larger long-term collaborations.

## 6.4. Aligning Target Conferences/Journals

A common pain point in cross-community collaboration is the difference in publication venues and tenure priorities. The conference-oriented culture of AI favors faster development and iterative improvement, whereas the journal-oriented culture of OR prioritizes less frequent standalone contributions. While both perspectives have merit, they create friction in project length/depth and target venues. We suggest the establishment of additional targeted research conferences jointly hosted by leading investigators from the two communities, following the model of existing conferences such as CPAIOR and the ACM Conference on Economics and Computation. These have emerged as forums where both research communities interact and exchange ideas on related research.

More broadly, communities can educate each other about high-quality publishing venues in their respective fields. Currently, advances in predictive and prescriptive analytics are each typically pursued by and celebrated in different communities. Better predictive models and algorithms are more likely to be investigated by and published in venues focused on AI scholarship (i.e., ML conferences); and better prescriptive models and algorithms tend to be investigated by and published in venues focused on OR scholarship (i.e., mathematical optimization journals). There are certainly opportunities at the intersection of both methodologies and their applications, which may end up celebrated and known in only one of those two communities, or perhaps not pursued at all due to a misalignment between the AI and OR communities in terms of expectations about the results and incentives for publications.

We want to ensure that promotion and tenure policies "count" publications in ways that are appropriate to the faculty member. In addition, there should be clear policies on how major OR journals will evaluate published conference papers at top ML conferences. We can consider models such as the ACM Conference on Economics and Computation, which accommodates the publication culture of both communities by allowing the publication of abstracts and gives the opportunity for accepted papers to be fast tracked for publication in Management Science, thus enabling and encouraging the submission of extended versions of conference papers through special issues and/or expediting the review process of such papers. By establishing a mutual understanding of reputable venues, we can increase alignment for potential collaborators and incentivize cross-publication between AI and OR/MS venues.

## 6.5. Benchmark Creation

Finally, we propose a concrete initiative to encourage joint work. We observe that the AI community has found great success in the extensive use of "benchmark" datasets to enable standard comparison of different algorithms on a common set of tasks. Likewise, the OR community has benchmarks focused on specific types of problems, some being more general such as MIPLIB, QPLIB, and MINLPLIB, and others more focused, such as TSPLIB, the many VRP benchmarks, and the classic OR-Library by J.E. Beasley. We believe that there is room for benchmarks involving problems in which both AI and OR are required for a complete approach.

Moreover, we would like to underscore the pressing need for complex and multi-layered benchmark datasets in contemporary research. New benchmarks must transcend simple, single-layer datasets (e.g., those that can be tackled by a single ML technique), and evolve into richer, more nuanced collections that encapsulate various dimensions of human and societal dynamics. This is vital for addressing and unraveling the intertwined layers of challenges that modern applications present, especially those that are socio-economically and systemically significant, such as those involving infrastructural systems (e.g., power system operations) and human interactions. A key example is the COMPAS dataset on predicting the risk of recidivism for defendants, which sparked widespread discourse on ethical AI by allowing researchers to explore issues around race, gender, and metrics of model performance, within a high-impact application context. The call is for benchmarks that not only provide a base for evaluating analytics models but also promote joint research by including data that reflect complex human interactions and systemic constraints. This multifaceted perspective on benchmarks will foster a more comprehensive understanding of the application domain and the societal narratives they weave, enhancing the relevance and impact of research in the real world. Such datasets will naturally benefit GenAI applications due to alignment on concrete use cases, domains, and problems to address, as well as the creation of appropriate datasets for training, testing and validation of these models. Given the rapid emergence of the field and the insufficiency of standard evaluation schemes, we believe this is an active area that would benefit from both communities' perspectives.

# ACKNOWLEDGMENTS

Reviewers CCC and the authors acknowledge the reviewers whose thoughtful comments improved the report:

## Final Report Reviewers

Daniel Lopresti, Lehigh University, United States

Anjana Susarla, Michigan State University, United States

# APPENDIX

## Workshop 1

September 23-24, 2021, held virtually

*Workshop 1 Organizers*

Sanmay Das, George Mason University

John Dickerson, University of Maryland

Pascal Van Hentenryck, Georgia Institute of Technology

Sven Koenig, University of Southern California

Ramayya Krishnan, Carnegie Mellon University

Radhika Kulkarni, SAS Institute, Inc. (retired)

Phebe Vayanos, University of Southern California

## Workshop 2

August 16-17, 2022 in Atlanta, GA

*Workshop 2 Organizers*

John Dickerson, University of Maryland

Bistra Dilkina, University of Southern California

Yu Ding, Georgia Institute of Technology

Swati Gupta, Massachusetts Institute of Technology

Pascal Van Hentenryck, Georgia Institute of Technology

Sven Koenig, University of Southern California

Ramayya Krishnan, Carnegie Mellon University

Radhika Kulkarni, SAS Institute, Inc. (retired)

## Workshop 3

March 21-22, 2024 in Washington, D.C.

*Workshop 3 Organizers*

[Yu Ding](#), Georgia Institute of Technology

[Phebe Vayanos](#), University of Southern California

[Sven Koenig](#), University of Southern California

[Ramayya Krishnan](#), Carnegie Mellon University

[Radhika Kulkarni](#), SAS Institute, Inc. (retired)

## AI/OR Workshop 3 Participants

| First Name | Last Name | Affiliation |
| --- | --- | --- |
| Laura | Albert | INFORMS |
| Arielle | Anderer | Cornell University, Johnson Graduate School of Management |
| Martin | Bichler | Technical University of Munich |
| Dan | Boley | University of Minnesota |
| Gianluca | Brero | Brown University |
| Cornelia | Caragea | National Science Foundation |
| Arthur | Choi | Kennesaw State University |
| Pooja | Dewan | Otis Elevators |
| John | Dickerson | University of Maryland |
| Bistra | Dilkina | University of Southern California |
| Yu | Ding | Georgia Institute of Technology |
| Sharmila | Duppala | University of Maryland |
| Ferdinando | Fioretto | University of Virginia |
| Michael | Fu | University of Maryland |
| Catherine | Gill | Computing Research Association |
| Amy | Greenwald | Brown University |
| Haley | Griffin | Computing Research Association |
| Swati | Gupta | Massachusetts Institute of Technology |
| Vishal | Gupta | University of Southern California Marshall School of Business |
| Philip | Keenan | General Motors |
| Sven | Koenig | University of Southern California |
| Ramayya | Krishnan | Carnegie Mellon University H. John Heinz II College |
| Radhika | Kulkarni | SAS Institute, Inc. (retired) |
| Connor | Lawless | Cornell University |
| Michael | Littman | National Science Foundation |
| Leqi | Liu | Princeton University / UT Austin |
| Andrea | Lodi | Cornell Tech and Technion - IIT |
| Lavanya | Marla | University of Illinois at Urbana-Champaign |
| Karthika | Mohan | Oregon State University |
| Thomy | Phan | University of Southern California |
| Amin | Rahimian | University of Pittsburgh |
| Xingyu | Ren | The University of Maryland, College Park |
| Nima | Safaei | Scotiabank |
| Thiago | Serra | Bucknell University |
| Sahil | Shikalgar | Northeastern University |
| David | Shmoys | Cornell University |
| Julie | Swann | North Carolina State University - Fitts Department of Industrial and Systems Engineering |
| Matthew | Titsworth | General Motors |
| Varun | Valsaraj | SAS Institute Inc |
| Pradeep | Varakantham | Singapore Management University |
| Phebe | Vayanos | University of Southern California |
| Segev | Wasserkrug | IBM |
| Holly | Wiberg | Carnegie Mellon University |
| Zhenyu | Yue | University Of Maryland, College Park |